\documentclass[aps,prb,twocolumn,superscriptaddress,showpacs,amsmath,amssymb]{revtex4}
\usepackage{graphicx}
\usepackage{natbib}
\usepackage{bm}

\newcommand{\LICUO}{LiCu$_2$O$_2$}

\begin{document}

\title{Spin waves and magnetic interactions in \LICUO.}

\author{T. Masuda}
\affiliation{Condensed Matter Sciences Division, Oak Ridge
National Laboratory, Oak Ridge, TN 37831-6393, USA.}

\author{A. Zheludev}
\affiliation{Condensed Matter Sciences Division, Oak Ridge
National Laboratory, Oak Ridge, TN 37831-6393, USA.}
\email{zheludevai@ornl.gov}
\homepage{http://neutron.ornl.gov/~zhelud/}

\author{B. Roessli}
\affiliation{Laboratory for Neutron Scattering, ETH Zurich and
Paul Scherrer Institute, CH-5232 Villigen PSI, Switzerland.}

\author{A. Bush}
\affiliation{Moscow Institute of Radiotechnics, Electronics and
Automation, Moscow 117464, Russia.}

\author{M. Markina}
\affiliation{Low Temperature Physics Department, Moscow State
University, Moscow 119992, Russia.}

\author{ A. Vasiliev}
\affiliation{Low Temperature Physics Department, Moscow State
University, Moscow 119992, Russia.}

\date{\today}
\begin{abstract}
The quasi-one-dimensional helimagnet \LICUO\  was studied by
single crystal inelastic neutron scattering. The dispersion
relation of spin wave excitations was measured in the vicinity of
the principal magnetic Bragg reflection. A spin wave theoretical
analysis of the data yields an estimate of the relevant exchange
constants and explains the mechanism of geometric frustration that
leads to helimagnetism. It is found that the simple
antiferromagnetic $J_1-J_2$ model that was previously proposed is
inadequate for \LICUO. The experimental findings are generally in
a qualitative agreement with first principal calculations of [A.
A. Gippius {\it et al.}, Phys. Rev. B {\bf 70}, 020406 (2004)],
though certain important discrepancies remain to be explained.

\end{abstract}

\pacs{}

\maketitle
\section{Introduction}

\LICUO\ is a quasi-one-dimensional magnetic material that has
recently attracted a great deal of
attention.\cite{Vorotynov,Fritschij98,Zvyagin,Zatsepin} Muon spin
resonance\cite{Roessli}, neutron scattering
experiments\cite{Masuda2004} and NMR studies\cite{Gippius2004}
clearly demonstrated that this $S=1/2$ system has an
incommensurate helimagnetic ground state. In our previous
work\cite{Masuda2004} we postulated a simple model of magnetic
interactions that seemed to account well for the observed
properties. It was proposed that \LICUO\ features weakly coupled
zig-zag $S=1/2$ chain with competing nearest-neighbor $J_1$ and
next-nearest-neighbor $J_2$ {\it antiferromagnetic} interactions.
Due to geometric frustration this model exhibits an unusual
competition between a quantum-disordered gapped\cite{nnn} ground
state with commensurate spin correlations,\cite{Chitra95} and
semi-classical helimagnetic incommensurate state. However,
independent first-principle numerical
studies\cite{Gippius2004,Drechsler2004} suggested a totally
different model for the geometric frustration in \LICUO. According
to that work, the nearest-neighbor $J_1$ interactions are actually
{\it ferromagnetic}, and it is the unusually large {\it
fourth}-nearest neighbor antiferromagnetic coupling constant $J_4$
that causes geometric frustration and favors helimagnetism. A
similar scenario was recently proposed for a very similar
isostructural helimagnet NaCu$_2$O$_2$.\cite{Capogna2004}

Bulk magnetic susceptibility data for \LICUO\ seem to be
compatible with both models,\cite{Masuda2004,Drechsler2004} and
are thus inconclusive. The only direct way to determine
(effective) exchange constants is by means of single crystal
inelastic neutron spectroscopy. This technique can probe the
dispersion relation of spin wave excitations in the system. In the
present paper we describe the results of such a study. We find
that the actual topology of magnetic coupling is indeed more
complex than originally envisioned.

\subsection{Structural considerations}
\begin{figure}
\includegraphics[width=3.3in]{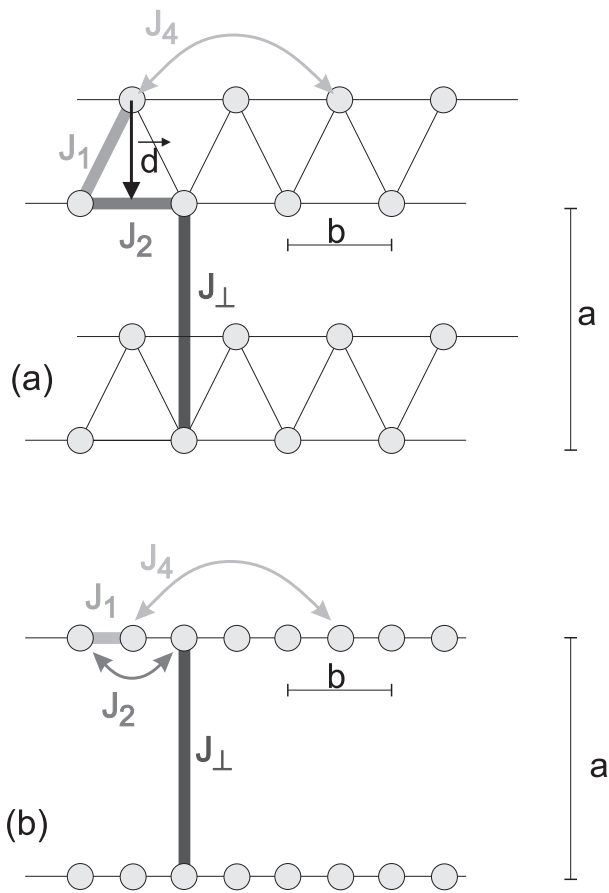}
 \caption{\label{coupling} (a) A schematic view of exchange interactions between magnetic Cu$^{2+}$
 ions in \LICUO. (b) An equivalent Bravais lattice of spins obtained by displacing every
 other Cu$^{2+}$ ion in the original non-Bravais lattice by the vector $\mathbf{d}$.
 It is assumed that the spin Hamiltonian remains intact upon this transformation. }
\end{figure}

\LICUO\ has a layered charge-ordered orthorhombic crystal
structure (space group Pnma, $a = 5.730(1)$\AA, $b =
2.8606(4)$\AA, and $c = 12.417(2)$\AA), as described in detail in
Refs.~\onlinecite{Zvyagin,Roessli}. The magnetism is due to double
chains of Cu$^{2+}$ ions that run along the crystallographic $b$
axis. The period of each ``leg'' of the double  spin chains is
equal to $b$. The two legs are offset by $b/2$ relative to each
other and thus form a ``triangular ladder'', as illustrated in
Fig.~\ref{coupling}a. These double chains are arranged in layers
parallel to the $(a,b)$ plane and are effectively separated along
the $c$ direction by planes of non-magnetic Cu$^{+}$ ions.

The approximate magnetic structure was determined in
Ref.~\onlinecite{Masuda2004} and is illustrated in Fig.~1 of that
paper. The ordering temperature is $T_\mathrm{c}=24$~K. Each
double spin chain carries a planar helix of spins. The magnetic
propagation vector is $\mathbf{q}_0=(0.5, \zeta,0)$ with
$\zeta\approx 0.827$. Consecutive spins on each rung are almost
parallel, being rotated relative to each other by an angle
$\phi=2\pi \zeta$. Within each double chain any nearest-neighbor
spins from opposite legs are almost antiparallel and form an angle
$\phi/2=\pi \zeta$. Preliminary neutron diffraction experiments
were consistent with $(a,b)$ being the spin rotation plane, but
independent NMR studies clearly show an out-of-plane
component.\cite{Gippius2004} This suggests that the spin rotation
planes are, in fact, tilted relative to a high-symmetry
orientation. The issue deserves further investigation, perhaps
involving more accurate diffraction measurements. However, if
magnetic anisotropy effects are negligible compared to Heisenberg
exchange interactions (a reasonable assumption for Cu$^{2+}$
ions), the actual arrangement of spin rotation planes should not
affect spin wave dispersion relations that are the focus of the
present study.

The main goal of the inelastic neutron scattering experiments
described below is to establish the hierarchy of exchange
interactions. The corresponding coupling constants are
schematically shown in Fig.~\ref{coupling}a, and define a model
Heisenberg Hamiltonian for \LICUO:
 \begin{eqnarray}
 \tilde{H}&=&\sum_{i,j} \left[ J_1\mathbf{S}_{i,j}\mathbf{S}_{i+1,j}
 +J_2\mathbf{S}_{i,j}\mathbf{S}_{i+2,j}+ \right. \nonumber \\
 &+& \left. J_4\mathbf{S}_{i,j}\mathbf{S}_{i+4,j}+J_\bot\mathbf{S}_{i,j}\mathbf{S}_{i,j+1}
 \right],\label{Ham}
 \end{eqnarray}
where the index $i$ labels consecutive spins in each double chain,
as shown in Fig.~\ref{coupling}a, and $j$ labels the double
chains.

\section{experimental}
In the present study we employed a 5~g single-crystal sample of
\LICUO. Crystal mosaic was not particularly good: irregular, with
as much as $4^\circ$ FWHM spread. This circumstance imposed
certain constraints as will be discussed below. The measurements
were carried out in two separate series of experiments. Neutrons
with a fixed final energy of $E_f=13.5$~meV were used at the HB-1
thermal 3-axis spectrometer at the High Flux Isotope Reactor at
ORNL (Setup 1). Pyrolitic graphite PG (002) reflections were used
for monochromator and analyzer. A PG filter was installed after
the sample to eliminate higher-order beam contamination. The setup
employed  $48'-80'-80'-240$ collimators. Alternatively, we
utilized cold neutrons with a final energy fixed at $E_f=5$ meV at
the TASP 3-axis instrument installed at the SINQ spallation
source, Villigen PSI, Switzerland (Setup 2). PG monochromator and
analyzer were used with no additional collimation and a cold Be
filter positioned after the sample. In all cases the sample was
mounted with the $(a,b)$ plane coinciding with the horizontal
scattering plane of the instrument. Sample environment was a
standard ``Orange''-type flow cryostat. Most of the data were
taken at $T=1.5$~K.

Data collection procedures were aimed at minimizing the adverse
effects of the broad mosaic spread of the sample. A large mosaic
translated to a poor wave vector resolution perpendicular to the
momentum transfer $\mathbf{q}$. Since the spin wave dispersion is
steepest along $b^\ast$, the best focusing conditions are obtained
for $\mathbf{q}$ pointing close to that direction. Considering the
intensity reduction due to the effect of magnetic form factors at
large $|q|$, an optimal ``window'' for low-energy spin wave
measurements was identified in the vicinity of the $(0.5,0.827,0)$
and $(0.5,1.173,0)$ magnetic Bragg peaks. Another advantage of
this reciprocal-space region is that it is relatively clear of
spurious scattering and phonons, that were carefully checked for.
Most scans were repeated at $T=75$~K, {\it i.e.}, well above
$T_\mathrm{c}$, to verify that the signal observed at low
temperature is indeed of magnetic origin. In some cases smoothed
high-temperature scans were used as background for point-by-point
subtraction from the corresponding low-temperature data sets.

Another important technical problem that had to be dealt with is
crystal twining. In \LICUO\ twinning occurs at a microscopic
level.\cite{Zvyagin} The $a$ axis of one type of domains coincides
with the $b$ axis of the other domain type, since $a\approx 2b$.
The immediate consequence is that neutron diffraction and
inelastic scattering necessarily detect a joint signal originating
from both domain types. As will be discussed in detail below, the
contributions from different domains and different spin wave
branches within each domain could be reliably separated only in
the direct proximity of the $(0.5,0.827,0)$ and $(0.5,1.173,0)$
peaks. Even so, twinning reduces the effective sample volume by a
factor of two. To avoid any confusion we will use the indexes $h$,
$k$ and $l$ to label wave vector components in the coordinate
system defined by the reciprocal lattice of one particular domain
type (type-A). Indexes $h_B=2k$, $k_B=h/2$ and $l_B=l$ will refer
to the other crystallographic domain (type-B).

\section{Results}

\begin{figure}
\includegraphics[width=3.3in]{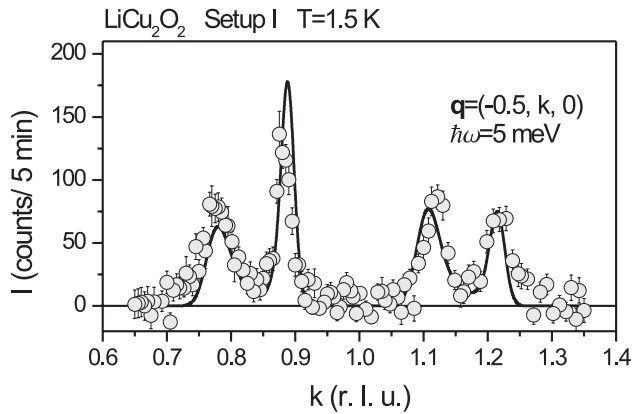}
 \caption{\label{exdata1} A constant-$E$ scan along the $(-0.5,k,0)$ reciprocal-space
 rod measured in \LICUO\ at $T=1.5$~K at an energy transfer of 5~meV (symbols) using Setup 1. The lines
 are a simulation based on the measured spin wave dispersion relation and the known
 resolution function of the instrument. }
\end{figure}

\begin{figure}
\includegraphics[width=3.3in]{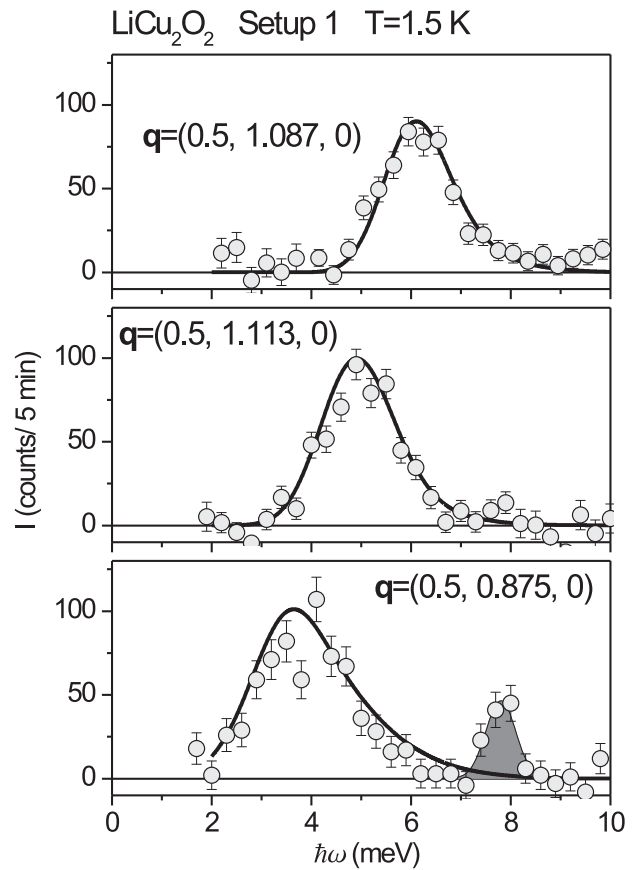}
 \caption{\label{exdata2} Typical constant-$q$ scans measured in \LICUO\ (symbols) at $T=1.5$~K.
 The lines are as in Fig.~\protect\ref{exdata1}. The shaded area in the lower scan is an
 ``accidental Bragg'' spurious peak originating from $2k_i$ scattering in the monochromator, $(0,2,0)$
 Bragg scattering in the sample and inelastic thermal-diffuse scattering in the analyzer. The spurious peak
 appears much narrower than the experimental resolution (about 2~meV FWHM). }
\end{figure}

\begin{figure}
\includegraphics[width=3.3in]{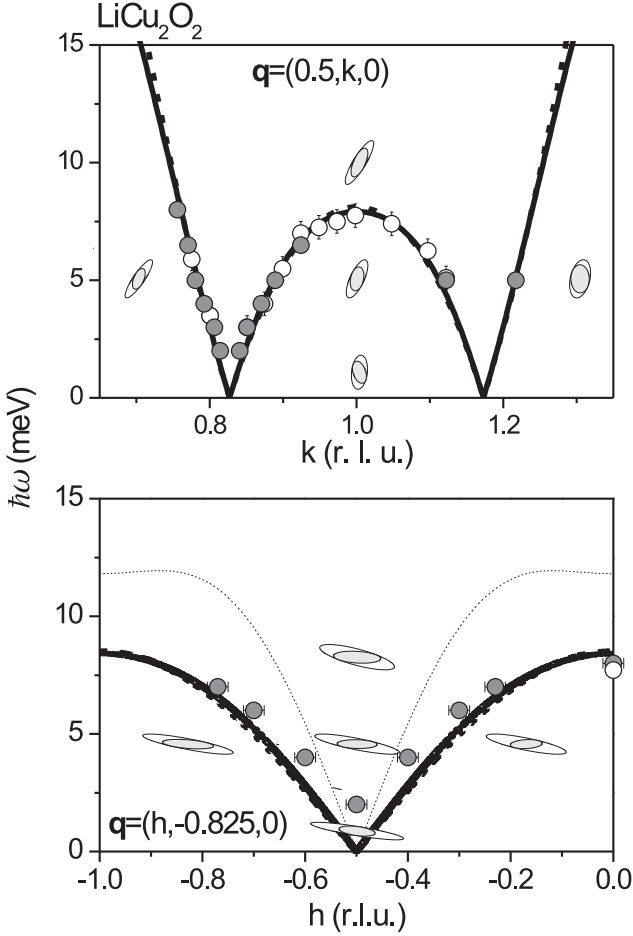}
 \caption{\label{disp} Spin wave dispersion measured in \LICUO\ at $T=1.5$~K (symbols). Shaded and
 open circles are data points obtained from constant-$E$ and constant-$q$ scans, respectively.
 The lines are a fit to the data, as described in the text. Heavy
 solid, heavy dashed and thin dotted lines correspond to Models 1,
 2 and 3, respectively. In both panels the ellipses represent the FWHM of
 the instrument resolution function in the appropriate projection (open ellipse) and section (shaded ellipse).}
\end{figure}

A typical constant-energy scan collected using Setup 1 is shown in
Fig.~\ref{exdata1}. The non-magnetic background for this scan
(featureless and typically about 80 counts per 5 min.) was
measured at $T=75$~K. A linear fit to the measured background was
subtracted from the data shown. The four prominent peaks seen in
this scan can be attributed to an acoustic spin wave emanating
from two magnetic Bragg peaks at $(0.5,0.827,0)$ and
$(0.5,1.173,0)$. Representative constant-$q$ scans collected using
the same setup are shown in Fig.~\ref{exdata2}, where a flat
background has been subtracted from the data. All inelastic peaks
observed for $0.75 \lesssim k \lesssim 1.25$ and $\hbar\omega
\lesssim 8$~meV can be associated with a single branch of
excitations.  The corresponding dispersion relation along the
crystallographic $b$ axis was extracted from the measured scans
using Gaussian fits. The result is plotted in circles  in
Fig.~\ref{disp}a, where solid and open symbols indicate
constant-$q$ and constant-$E$ measurements, respectively. Any
attempts to follow the observed spin wave branch to higher
energies were not successful. The mode's intensity drops
progressively, while the background  increases and becomes
structured. As will be explained below, part of this problem may
be due to a multitude of additional spin wave branches from both
types of domains. The situation was further aggravated by limited
resolution, phonon scattering and instrument-related spurious
peaks. For this reason, throughout this paper we shall limit the
discussion to experimental data collected at low energies in
direct proximity of the two above-mentioned magnetic Bragg peaks.

 Additional measurements were performed using
Setup 2 to determine the dispersion along the $a^\ast$ axis.
Typical scans and the measured dispersion curve are shown in
Figs.~\ref{exdata3} and \ref{disp}b, respectively. The important
zone-boundary scan at $(0,0.827,0)$ was also measured using Setup
1 in the range 2--15~meV and found to be fully consistent with
that shown in Fig.~\ref{exdata3}c.

\subsection*{Data analysis} In order to extract the relevant
exchange constants from the measured dispersion curves the data
were analyzed in the framework of semiclassical spin wave theory
(SWT).\cite{Nagamiya}

\subsubsection{Equivalent Bravais lattice} At a first glance,
calculating the dispersion relation is quite tedious, since we are
dealing with a non-Bravais lattice of spins with four magnetic
ions per unit cell. Within the approximation of the Hamiltonian
(\ref{Ham}), however, any interactions between the double-chain
layers are ignored, and only two spins per unit cell remain.
Moreover, the spin network can be made equivalent to one on a
Bravais lattice by displacing the atoms as shown in
Fig.~\ref{coupling}b. The dynamic structure factor
$S_0(\mathbf{q},\omega)$ of this ``straightened out'' spin network
is directly related to that of the original model through:
 \begin{subequations}
 \begin{eqnarray}
 S(\mathbf{q},\omega)&=&S_0(\mathbf{q},\omega)\cos^2\frac{\mathbf{qd}}{2}
 +\nonumber\\
 &+&
 S_0(\mathbf{q}+(0,1,0),\omega)\sin^2\frac{\mathbf{qd}}{2},\label{zigzagterms}
 \end{eqnarray}
 \begin{equation}
 \mathbf{qd}=2\pi h \delta_x+2\pi l \delta_z,
 \end{equation}
 \label{zigzag}
 \end{subequations}
where $\mathbf{q}=(h,k,l)$ is the wave vector transfer and
$\mathbf{d}$ is the relative displacement of the two legs in each
double chain: $\mathbf{d}=\mathbf{a} \delta_x +
\mathbf{c}\delta_z$. For the crystal structure of \LICUO\ we have
$\delta_x=0.24$ and $\delta_z=0.19$. It is important to emphasize
that Eq.~\ref{zigzag} is {\it exact}: it does not imply any
particular properties of $S^{(0)}(\mathbf{q},\omega)$ and
$S(\mathbf{q},\omega)$, and relies {\it only} on the definitions
of these correlation functions.

\begin{figure}
\includegraphics[width=3.3in]{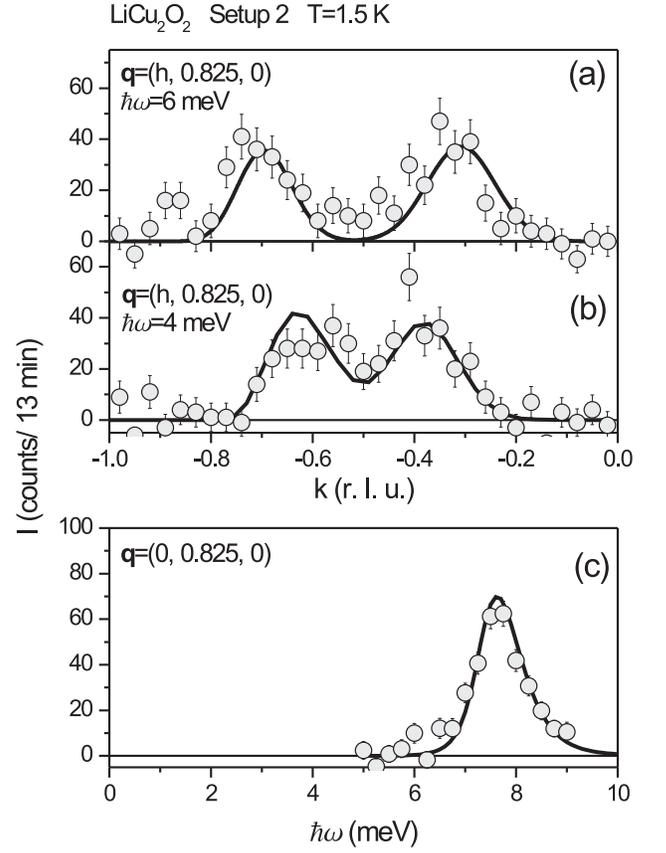}
 \caption{\label{exdata3} Typical scans
 measured in \LICUO\ using Setup 2 (symbols).
 The lines are as in Fig.~\protect\ref{exdata1}. }
\end{figure}

\begin{table*}
 \caption{\label{params} Sets of exchange constants obtained by fitting the
 calculated spin wave dispersion relation to the experimental curves in comparison
 with first-principle calculations.}
 \begin{ruledtabular}
 \begin{tabular}{c | l l l | l}
 & Model 1 & Model 2 & Model 3 &LDA\protect\footnote{From Ref.~\protect\onlinecite{Gippius2004}} \\
 \hline
  $J_1$ & 6.4(0.95)~meV & 105.5 (8.0)~meV & 0 (fixed) & 0.4~meV \\
  $J_2$\footnote{Not refined: fixed by Eq.~\protect\ref{con}} & -11.9~meV & 33.8~meV & -14.0~meV  & -8.1~meV\\
  $J_4$ & 7.4 (0.5)~meV & -1.6 (0.2)~meV & 7.5 (0.1)~meV & 14.4~ meV\\
  $J_\bot$ & 1.8(0.2)~meV & 0.23 (0.02)~meV & 6.8 (0.4)~meV&
  5.7~meV
 \end{tabular}
 \end{ruledtabular}
 \end{table*}

\subsubsection{SWT dynamic structure factor}

A very useful recipe for calculating the SWT dynamic structure
factor of a simple helimagnet on a Bravais lattice can be found in
Ref.~\onlinecite{Coldea2003}. The magnon dispersion relation is
given by:
\begin{subequations}
 \begin{eqnarray}
 \omega_\mathbf{q}^2 &=& A_\mathbf{q}^2-B_\mathbf{q}^2,\label{dispeq}\\
 A_\mathbf{q} &=& 2S\{
 \frac{J_\mathbf{q}}{2}+\frac{1}{4}\left[J_{\mathbf{q}-\mathbf{q}_0}+J_{\mathbf{q}+\mathbf{q}_0}\right]-J_{\mathbf{q}_0}\},\\
 B_\mathbf{q} &=& 2S\{
 \frac{J_{\mathbf{q}}}{2}-\frac{1}{4}\left[J_{\mathbf{q}-\mathbf{q}_0}+J_{\mathbf{q}+\mathbf{q}_0}\right]\},
 \end{eqnarray}
\end{subequations}
where, for our model Hamiltonian,
 \begin{eqnarray}
  J_{h,k,l}&=& J_1\cos(\pi k)+J_2\cos(2\pi k)+\nonumber \\
 & + &  J_4\cos(4\pi k)+J_\bot \cos(2\pi h).
 \end{eqnarray}
The requirement that $\mathbf{q}_0$ minimizes $J(\mathbf{q})$
fixes the relation between $J_1$, $J_2$, $J_4$ and $\zeta$:
\begin{equation}
J_1+4J_2\cos(\pi\zeta)+16J_4\cos(\pi\zeta)\cos(2\pi\zeta)=0.\label{con}
\end{equation}
Conveniently, $J_\bot$ does not enter this expression. The
single-magnon cross section has three contributions:
\begin{subequations}
\label{SQW} \begin{eqnarray}
 S^{\bot}_0(\mathbf{q},\omega)&=&\frac{S}{2}\frac{A_\mathbf{q}+B_\mathbf{q}}{\omega_\mathbf{q}}\delta(\omega-\omega_\mathbf{q}),\label{SQWz}\\
 S^{-}_0(\mathbf{q},\omega)&=& \frac{S}{8} \frac{A_{\mathbf{q}-\mathbf{q}_0}-B_{\mathbf{q}-\mathbf{q}_0}}{\omega_{\mathbf{q}-\mathbf{q}_0}} \delta(\omega-\omega_{\mathbf{q}-\mathbf{q}_0}),\\
 S^{+}_0(\mathbf{q},\omega)&=& \frac{S}{8}\frac{A_{\mathbf{q}+\mathbf{q}_0}-B_{\mathbf{q}+\mathbf{q}_0}}{\omega_{\mathbf{q}+\mathbf{q}_0}}\delta(\omega-\omega_{\mathbf{q}+\mathbf{q}_0}).
 \end{eqnarray}
\end{subequations}
Here $S^{\bot}(\mathbf{q},\omega)$ represents fluctuations of the
spin component perpendicular to the plane of the helix,  while
$S^{+}(\mathbf{q},\omega)$ and $S^{-}(\mathbf{q},\omega)$ are
in-plane excitations. The total SWT cross section for \LICUO\ is
related to $S_0(\mathbf{q}, \omega)$ through Eq.~\ref{zigzag}.

It now becomes clear why measuring spin waves in \LICUO\ is so
technically challenging: there are six spin wave branches in each
crystallographic domain. As a result, in any experiment one has to
deal with a total of {\it twelve} spin wave branches, all of which
are acoustic and therefore contribute to scattering at low
energies. Also, compared to a simple antiferromagnet on a Bravais
lattice, the scattering by each individual branch is typically
much weaker, since intensity is effectively re-distributed between
twelve modes. The existence of a suitable, albeit narrow,
measurement window that was exploited in our experiments  is a
fortunate coincidence.

\subsection*{Fits to experimental data}
The energy of the spin wave branch observed in our experiments
clearly goes to zero at the principal magnetic Bragg peaks of the
A-type domain. It must therefore be associated with the
$S^{\bot}(\mathbf{q},\omega)$ mode and the first term in
Eq.~\ref{zigzagterms}. An analysis of the polarization of spin
wave excitations is beyond the scope of this work. Nevertheless,
we can point out that $(a,b)$ being the spin rotation
plane\cite{Masuda2004} is consistent with a strong
$S^{\bot}(\mathbf{q},\omega)$. Indeed, the corresponding
polarization factor for unpolarized neutrons ia a maximum for
in-plane momentum transfers.\cite{Coldea2003}  As a first step in
our analysis we ignored all other spin wave branches and used
Eq.~\ref{dispeq} to fit the experimental data. It is easy to show
that there are exactly two sets of exchange parameters that {\it
exactly} reproduce (i) the experimentally determined
incommensurability parameter $\zeta$, (ii) the excitation energies
at the zone boundaries $(0.5,1,0)$ and $(0,0.827,0)$, and (iii)
the spin wave velocity along the $b^\ast$ direction near the
magnetic Bragg peaks. Using these two sets of parameters as
initial points, we employed a least-squares algorithm to best-fit
the data under the rigid constraint set by Eq.~\ref{con}. The two
resulting optimized sets of parameters are listed in
Table~\ref{params} and will be referred to as Models 1 and 2,
respectively. For a purpose that will be made clear in the next
section, we also attempted a fit to the data taken {\it only along
the $b$-axis} while fixing $J_1\equiv 0$. The measured $a$-axis
dispersion was not included in this fit. The resulting set of
exchange constants is also listed in Table~\ref{params} and will
be referred to as Model 3. Dispersion relations for the
out-of-plane branch and the first term in Eq.~\ref{zigzagterms}
calculated for type-A domains using Models 1 through 3 are plotted
in lines in Fig.~\ref{disp}.

\begin{figure}
\includegraphics[width=3.3in]{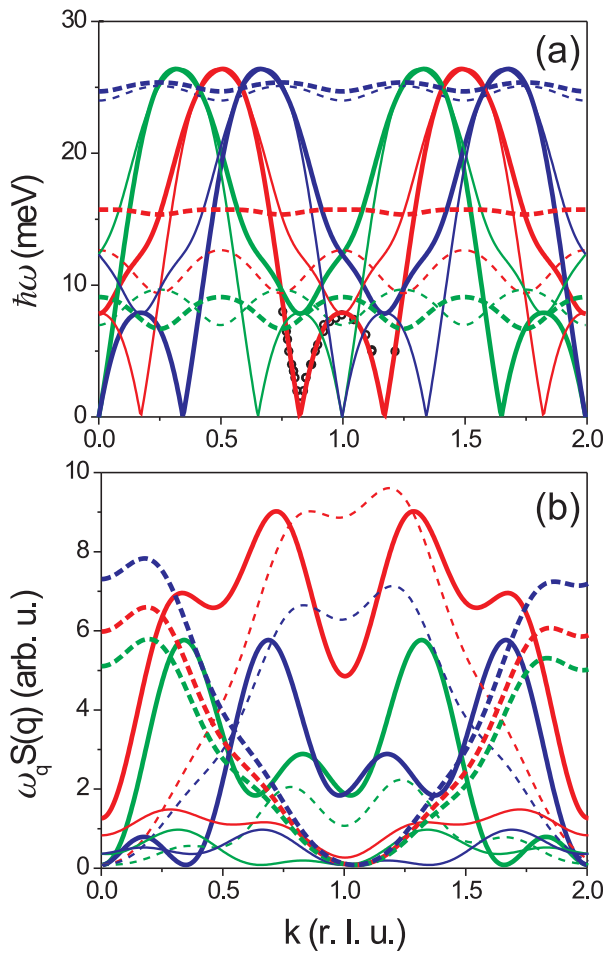}
 \caption{\label{alldisp} (a) Dispersion of all twelve spin wave branches along the $(0.5,k,0)$ reciprocal-space rod
 in twinned \LICUO\ crystals calculated using Model 1. (b) The corresponding structure factors scaled by energy transfer.
 In both panels the out-of-plane modes are plotted in red. Modes shown in blue and green lines are polarized in
 the plane of spin rotation. Thick and thin lines of all types correspond to the first and second terms in Eq.~\protect\ref{zigzagterms},
 respectively. Solid and dashed lines refer to type-A and type-B crystallographic domains. Symbols are as in
 the top panel of Fig.~\protect\ref{disp}.}
\end{figure}

The next step is to understand why the other 11 spin wave branches
were not observed in the studied wave vector and energy ranges.
All twelve dispersion curves calculated using Model 1 are plotted
in Fig.~\ref{alldisp}a.  The intensities of each branch were
scaled by the energy transfer (to get rid of the $1/\omega$
factors in Eq.~\ref{SQW}) and are plotted in Fig.~\ref{alldisp}b.
The neutron polarization factor was not included in this
calculation, and the maximum value (unity) was assumed for each
branch. One readily sees that all branches originating from type-B
domains and the 1st term in Eq.~\ref{zigzagterms} are extremely
weak in our area of interest. The same is true for all spin waves
associated with type-A domains and the 2nd term in
Eq.~\ref{zigzagterms}. Contributions from type-B domains and the
2nd term in Eq.~\ref{zigzagterms} are either weak or at very high
energy. In the end, only the ``principal'' modes originating from
type-B domains and the 1st term in Eq.~\ref{zigzagterms} are
relevant to our experiments. Of these three branches the two
in-plane modes are almost entirely outside the studied energy
range and do not affect any of the scans collected. As a result,
only one out-of-plane mode is seen. Similar arguments can be made
regarding the dispersion measured along the $a$-axis. The
simulation also explains why we were unable to follow the
``principal'' out-of-plane mode to higher energies. At around
8~meV energy transfer any inelastic scans become crowded with a
host of peaks associated with other spin wave branches that can
not be reliably resolved due to limited experimental resolution.

From Fig.~\ref{disp} it is apparent that the studied energy and
wave vector transfer range the dispersion relations produced by
Model 2 are virtually identical to those resulting from Model 1.
Both models fit the available experimental data rather well. The
only small discrepancy is seen along the $a$-direction where the
observed spin wave velocity near $\mathbf{q}=(-0.5,0.827,0)$
appears slightly larger than calculated, while the zone-boundary
energy at $\mathbf{q}=(0,0.827,0)$ is slightly smaller. This
discrepancy is easily explained if one assumes a small
anisotropy-induced gap $\Delta$ in the spin wave spectrum at the
magnetic Bragg peak position. From the existing scans it is
possible to tentatively estimate $\Delta\approx 1.5$~meV, though
more high-resolution data will be needed for a reliable
determination. As far as only the $b$-axis dispersion is
concerned, Model 3 seems to work just as well as the other two
sets of parameters, and for the $(0.5,k,0)$ reciprocal-space rod
produces almost identical curves. However, this good agreement
under the constraint $J_1=0$ {\it necessarily} results in a very
poor agreement with the $a$-axis dispersion relation. The
corresponding bandwidth calculated using Model 3 (thin dotted line
in Fig.~\ref{disp}b) is considerably larger than the observed
value. In other words, assuming $J_1=0$ in Hamiltonian \ref{Ham}
can not be made consistent with the bulk of the experimental data.

The resulting parameter values for Model 1 were used to simulate
the measured inelastic neutron scans. For this purpose we utilized
the magnetic dynamic structure factor as given by Eq.~\ref{SQWz}.
This model cross section was scaled by the magnetic form factor
for Cu$^{2+}$ and numerically convoluted with the 4-dimensional
spectrometer resolution function. The latter was calculated in
Gaussian form using the Cooper-Nathans approximation. The
evolution of the corresponding FWHM resolution ellipsoid in the
surveyed areas of $(E,\mathbf{q})$ space is shown in
Fig.~\ref{disp}. An overall scaling factor was chosen to best-fit
the measured scans. The results of such simulations are shown in
solid lines in Figs.~\ref{exdata1}--\ref{exdata3}. Overall, the
experimental peak shapes are well explained by resolution effects
alone.

\section{Discussion}
As mentioned in the introduction, the knowledge of spin wave
dispersion in an ordered magnetic material can, in principle,
provide comprehensive information on the exchange interactions.
This requires measurements of dispersion curves across the entire
Brillouin zone. In the particular case of \LICUO, twinning, less
than perfect sample mosaic, and a complex spectrum limit us to
only observing spin waves in the direct proximity of the principal
magnetic Bragg reflections. For this reason the resulting sets of
parameters can be regarded as unique only under the assumption of
the validity of Hamiltonian \ref{Ham} and the SWT approximation.

The Hamiltonian has a solid justification in the crystal
structure, and is well supported by first-principle calculations.
However, a vital question is to what extent an analysis based on
essentially classical spin wave theory can be valid in the case of
a quasi-low-dimensional frustrated magnet such as \LICUO? At the
very least, the exchange constants obtained from such an analysis
will be renormalized compared to their actual values due to
quantum corrections. Nevertheless, even for highly anisotropic or
frustrated systems, acoustic  modes can be associated with waves
in the order parameter field. In this case a {\it renormalized}
SWT with some {\it effective} set of exchange parameters typically
works rather well. Just two examples are the weakly-coupled
$S=1/2$ chains compound KCuF$_3$ (Ref.~\onlinecite{Tennant95}) and
the quasi-2D frustrated helimagnet
Cs$_2$CuCl$_4$.\cite{Coldea2003} Deviations from classical
behavior in \LICUO\  do not appear particularly severe. The
typical energy scale of exchange interactions computed by LDA
exceeds the temperature of 3D long-range ordering by less than an
order of magnitude. While the absolute value of the ordered moment
has not been determined to date, that in the very similar
helimagnet NaCu$_2$O$_2$ is quite large ($0.64~\mu_\mathrm{B}$
).\cite{Capogna2004} We therefore have good reason to assume that
{\it renormalized} SWT will work well for our system.

Turning now to the two solutions that we obtained we note that
Model 2 almost exactly corresponds to the simple $J_1-J_2$ model
that we originally proposed.\cite{Masuda2004} However, unless the
renormalization of exchange parameters is severe and {\it all}
actual in-chain exchange constants are much smaller than the
effective SWT ones, the estimated parameters are not consistent
with a previous analysis of magnetic
susceptibility.\cite{Masuda2004} We thus conclude that Model 1 is
a much more likely to be an adequate description of \LICUO. It's
key features, namely a {\it ferromagnetic} $J_2$ bond and a
substantial $J_4$ coupling constant, are similar to those of the
LDA calculations of Ref.~\onlinecite{Gippius2004}. Geometric
frustration in \LICUO\ is thus similar to that in
NaCu$_2$O$_2$.\cite{Capogna2004}

The main difference between the exchange parameters in Model 1 and
those emerging from LDA calculations of
Ref.~\onlinecite{Gippius2004} is in the magnitude of $J_1$. As
illustrated by Model 3, assuming that $J_1$ is negligibly small
makes the SWT calculation incompatible with the measurements. We
conclude that either LDA severely underestimate $J_1$, or this
exchange constant is renormalized by over an order of magnitude in
the SWT. Of course, one can never entirely rule out the
possibility that both the LDA calculations and our SWT analysis
are missing some crucial terms in the Hamiltonian. Further
experimental and theoretical insight will be required to fully
resolve this remaining mystery.

\section{Conclusion}
In summary, based on inelastic neutron scattering measurements we
can conclude that the simple $J_1$--$J_2$ model that we originally
proposed does not apply to \LICUO. The frustration mechanism is
more complex and resembles that proposed in
Refs.~\onlinecite{Gippius2004,Drechsler2004}. It involves a
competition between a combination of antiferromagnetic $J_1$ and
ferromagnetic $J_2$ interactions against an additional
antiferromagnetic long-range $J_4$ coupling. We find the three
corresponding exchange constants to be of comparable absolute
strength. This discrepancy between an SWT-based interpretation of
the experimental data and first-principle calculations remains to
be explained.

\acknowledgements One of the authors (AZ) would like to thank A.
Drechsler for numerous insightful discussions on the subject and
for his constructive comments\cite{Drechsler2004} on
Ref.~\cite{Masuda2004}. This work was partially supported by the
Civilian Research and Development Foundataion project
RU-P1-2599-04. Work at ORNL was carried out under Contracts No.
DE-AC05-00OR22725, US Department of Energy. Part of this work was
done at the Swiss Neutron Spallation Source SINQ, Switzerland.


\end{document}